# Global YouTube Trending Dataset (2022-2025): Three Years of Platform-Curated, Cross-National Trends in Digital Culture

Alexandre Goncalves[1], Yee Man Margaret Ng[1]

[1]University of Illinois Urbana-Champaign
gonca@illinois.edu, ymn@illinois.edu

**Abstract**

On July 1, 2025, YouTube retired its decade-long public "Trending" pages, ending platform-curated, non-personalized video discovery. The Trending list had long served as a vital lens into algorithmic influence, cultural diffusion, and crisis communication globally, offering a rare "ground-truth" reference to study global attention and cultural salience. We present a three-year archival dataset of YouTube Trending videos, collected from July 1, 2022, to June 30, 2025, with four daily snapshots for each of the 104 countries. The dataset includes 446,971 snapshots, each capturing up to 200 trending videos, encompassing 78.4 million video entries (726,627 unique videos) and associated metadata. Each record includes core identifiers (snapshot time, country, rank) and content metadata (video ID, channel ID, title, description, tags, publication date, category, channel name, language, live status, views, and comments). Unlike previous datasets with limited geographic scope or short timeframes, our non-personalized data provides exceptional cross-national and longitudinal coverage for studying digital culture, platform governance, and temporal dynamics in content popularity. We document the data collection methodology, schema design, coverage, descriptive statistics for both global and U.S. trending videos, and the ethical safeguards implemented throughout.

**Datasets** — https://doi.org/10.13012/B2IDB-9307654_V1

## 1 Introduction

YouTube plays a central role in today's media landscape. With over 2 billion monthly users watching more than 1 billion hours each day, it functions as a venue for entertainment, information, news, and public debate (Burgess and Green 2018; YouTube 2024). Its features—multimodal formats (visual, audio, text), low barriers to entry (anyone can upload), and social tools (subscriptions, comments, sharing)—have made it a key site for studying cultural circulation (Ng and Taneja 2023), audience formation (Yang et al. 2022), and shifts in information salience across time and countries (Spartz et al. 2015).

A video's impact is usually measured by both its long-term popularity (lifetime views) and short-term visibility (trending status). In YouTube, while lifetime views show eventual reach, the platform also highlighted momentary visibility through its "Trending" page. For over a decade, this page displayed videos that were "hot" in specific countries. The selection criteria extended beyond total views to include factors such as video topic and age, along with both the rate and acceleration of view growth (YouTube Help 2020). The YouTube Trending page had long been regarded as an active form of platform intervention: a curated list that reflected not only audience interest but also platform choices about what to promote. Scholars viewed this list as an instance of algorithmic agenda-setting, shaping collective attention and cultural salience (Bucher 2018; McCombs and Shaw 1972).

However, YouTube officially discontinued its "one all-encompassing" Trending list in July 2025, ending a feature that defined popular content in different regions for nearly a decade (Malik 2025; YouTube Help 2025). This strategic move toward category-specific charts (music videos, podcast shows, and movie trailers) reflects a broader shift toward algorithm-driven personalization, where individual user preferences supplant universal benchmarks of popularity (Marketing Tech 2025). This change may be a direct consequence of an industry-wide move toward hyper-personalization, which maximizes watch time and builds stronger creator-viewer relationships to achieve platform profitability. By contrast, the Trending page had functioned as a non-commercial "public good." Its declining engagement meant its maintenance was no longer aligned with evolving business objectives.

The need for platform-level visibility data has grown amid concerns about misinformation, polarization, and algorithmic amplification. YouTube's recommendation algorithms have become central to this debate, with critics arguing they create filter bubbles, drive radicalization, and deepen political divides. The findings on this phenomenon are not uniform; while some studies argue algorithms direct users toward problematic content (Basch et al. 2020; Li et al. 2020), others dispute that user choice is a more significant factor (Ribeiro et al. 2020). This debate is often hindered by a lack of "ground truth" data—a neutral corpus of popular content not shaped by individual user history or preferences (Yesilada and Lewandowsky 2022). The removal of the Trending page removes one of the few non-personalized reference points available for such work.

Furthermore, longitudinal, multi-country trending data is scarce. Most YouTube datasets are limited in scope, focusing on single countries (Brodersen, Scellato, and

Wattenhofer 2012), short time periods (Faddoul, Chaslot, and Farid 2020), or specific channels (Ribeiro et al. 2020). To address this gap, this paper presents a three-year archival, longitudinal dataset of YouTube Trending videos from 104 countries, collected from July 1, 2022, to June 30, 2025. This dataset captures the final years of this public-facing discovery feature, rendering it an essential historical record for future studies of digital culture and communication. The non-personalized nature of this dataset allows researchers to disentangle the effects of algorithmic bias from general audience interest, addressing a core methodological challenge noted in existing literature.

## 2 Literature Review

### 2.1 YouTube as Information Infrastructure

YouTube transcends entertainment to function as a critical information infrastructure. Research documents its role in political mobilization (Tufekci 2017) and news consumption, with 26% of U.S. adults regularly getting news from the platform (Pew Research Center 2024). Its recommendation algorithm, processing over 80 billion signals daily (YouTube's Official Blog 2021), fundamentally shapes information exposure for billions (Zhao et al. 2019).

Studies of "convergence culture" show how YouTube blurs boundaries between professional and amateur media (Jenkins 2006; Cunningham and Craig 2019). However, despite narratives of global connectivity, consumption patterns remain strongly regionalized by language, culture, and infrastructure (Ng and Taneja 2023). This tension between a global platform and local practice makes cross-national trending data particularly valuable.

### 2.2 Algorithmic Agenda-Setting and Platform Power

Agenda-setting theory, foundational to communication research, argues media visibility shapes perceived salience of public agenda (McCombs and Shaw 1972). On algorithmic platforms, this influence is amplified by automated, personalized, and opaque curation (Napoli 2014). YouTube's Trending page functioned as an explicit form of algorithmic agenda-setting—a constantly refreshed feed of platform-determined national priorities (Rieder et al. 2018). Unlike personalized feeds, which are often studied through "algorithmic audits" (Sandvig et al. 2014), the Trending page provides a form of "ground-truth" visibility data. Research shows that trending videos experience 50-100x view amplification compared to non-trending content (Barjasteh et al. 2014), making inclusion consequential for both creators and audiences. Trending videos also exhibit "rich-get-richer" dynamics, with established channels dominating placement (Arthurs et al. 2018).

### 2.3 Platform Governance and Content Moderation

YouTube's content policies shape information access for billions, yet enforcement is often opaque (Kaye 2019; Gillespie 2018). Studies have documented bias concerns, including underrepresentation of female creators (Döring & Mohseni 2019) and overrepresentation of mainstream media (Arthurs et al. 2018; Burgess and Green 2018).

During the COVID-19 pandemic, concerns intensified over YouTube's dual role as both a key health information source and a channel for misinformation. Despite the platform's introduction of "authoritative source" interventions to reduce misinformation (YouTube 2020), Li et al. (2020) found that over a quarter of the most-viewed COVID-19 videos contained misleading claims, and Basch et al. (2020) reported inconsistent or inaccurate health guidance. However, broader algorithmic impacts remain debated: Hosseinmardi et al. (2021) found limited evidence of algorithmic bias, while Ng et al. (2023) argued that YouTube's promotion strategy reflected a "blockbuster" model that prioritized reliably successful content over ideological extremity.

Taken together, these results highlight the need for longitudinal, cross-national baselines. Evaluating platform governance and public-health communication requires systematic visibility data: what content platforms elevated before, during, and after policy interventions (Cinelli et al. 2020). Our dataset provides exactly that baseline by capturing the content YouTube elevated via its Trending page across these critical periods surrounding policy interventions.

## 3 Research Gaps and Dataset Value

YouTube Trends captures the platform-curated trending surface rather than relying solely on channel crawls. Much prior work relies on channel-based collections that reflect what users upload or subscribe to. Trending, by contrast, is platform-selected and reveals what YouTube itself highlights as nationally relevant. Studying this surface is crucial for questions of agenda setting and platform governance, where the focus is on what the platform elevates (Barjasteh, Liu, and Radha 2014).

Most existing studies of YouTube Trending are limited to short-term or single-country samples. Our dataset overcomes these constraints, offering a resource for comparative, longitudinal, and event-driven analyses at a global scale.

### 3.1 Temporal Breadth and Longitudinal Scope

Our data spans the final three years of the Trending page (July 1, 2022-June 30, 2025), enabling analysis of platform evolution, shifting creator strategies, and attention cycles that short-term data cannot reveal.

## 3.2 High Sampling Frequency

Monthly or weekly snapshots overlook rapid changes: events and viral spikes can rise and fall within hours (Crane and Sornette 2008). With four captures per day, our dataset preserves ephemeral phenomena essential to crisis communication and viral-diffusion studies.

## 3.3 Cross-national Comparability

Research is frequently limited to single countries (Brodersen, Scellato, and Wattenhofer 2012), constraining comparative work on cultural diffusion, localization, and global-local dynamics. Our 104-country coverage supports tests of theories such as cultural proximity (Straubhaar 1991), linguistic markets (Wildman and Siwek 1988), and digital divides (van Dijk 2020), and allows empirical study of whether global platforms encourage convergence or reinforce regional differences (Ng and Taneja 2023; Webster and Ksiazek 2012).

## 3.4 Rich Multimodal Pairing

The dataset includes standard metadata, such as language fields, category labels, and broadcast status (live or recorded). It allows researchers to combine content analysis (titles, transcripts, tags) with audience engagement metrics (view and comment counts). Through the provided video IDs, researchers can retrieve additional data including full videos (visual and audio), transcripts, and viewer comments. Together, these resources support large-scale analyses of sentiment, toxicity, framing, and other content-audience interactions.

# 4 Dataset Description

## 4.1 Collection Protocol

For a decade, YouTube maintained lists of "Trending videos," which provide a rolling selection of the most popular content at a given moment and country. While the platform does not disclose user-level or country-level viewing statistics of a video, these Trending lists offered a rare, standardized snapshot of what was visible across regions. Using the YouTube Data API v3 (parameters for the request: *chart = mostPopular* and *regionCode*), we collected up to 200 trending videos per country at four fixed intervals each day (00:00-05:59, 06:00-11:59, 12:00-17:59, and 18:00-23:59 UTC). The monitoring ran continuously for three years, covering 104 countries between July 1, 2022, and June 30, 2025.

## 4.2 Storage

The raw data has been stored in Amazon S3 and made queryable through AWS Athena. Each retrieved record contains detailed information about the video, including identifiers (video ID, channel ID, and category ID), textual content (title and description), popularity metrics (view and comment counts), and localization fields (video and audio language).

## 4.3 Schema Overview

*most_popular.csv*: The main table that contains the primary data for trending videos.

**Collection Info**
- **collection_date:** Snapshot time (UTC)
- **region_code:** Country code where the video trended
- **rank:** Position in the trending list

**Content & Metadata**
- **video_id:** Unique video ID
- **channel_id:** Unique channel ID
- **title:** Video title
- **description:** Text description
- **published_at:** Date and time when video was published
- **channel_title:** Channel display name
- **category_id:** Video category ID
- **default_language:** Metadata language (detected)
- **default_audio_language:** Audio language (detected)
- **live_broadcast_content:** Live broadcast vs. pre-recorded
- **view_count:** Number of views at collection
- **comment_count:** Number of comments at collection

*tags.csv*: Trending videos often have multiple tags assigned by their creators. Since CSV files cannot handle hierarchical or nested data structures, we store tags in a separate file rather than trying to fit multiple tags into a single field. To connect the tags with their corresponding videos, you can join this file with most_popular.csv using three matching Collection Info fields as a composite key (collection_date, region_code, rank).

**Collection Info**
- **collection_date:** Snapshot time (UTC)
- **region_code:** Country code where the video trended
- **rank:** Position in the trending list

**Content**
- **tag**: A single descriptive tag. Assigned by the content-creator.

## 4.4 Coverage and Quality

The complete dataset is approximately 26.4 GB in compressed format (tar.bz2). The dataset comprises 78,391,390 trending video entries across 446,971 snapshots. This represents a 98% coverage rate of the targeted 455,936 possible snapshots (4 times daily x 104 countries x 1,096 days), with most missing data usually due to brief system outages.

For the U.S., of the expected 4,384 snapshots (1,096 days x 4 daily retrieval), only 22 are missing, yielding 99.5% coverage. Sixteen of missing snapshots occurred during a brief system outage from 2022-12-01 00:00:00 to 2022-12-04 18:00:00, while the remaining six are isolated one-off gaps scattered throughout the collection.

## 5 Descriptive Statistics

To provide an overview of the dataset, we report several statistics that illustrate its scale, composition, and temporal patterns in both global and U.S. context.

### 5.1 Global Trend Overview

**Trends and repeats.** Out of 78.4 million trending video entries, 726,627 are unique videos. This hints recurrent visibility, algorithmic amplification, and cross-national diffusion.

**First appearance engagement**. When videos first appeared on YouTube Trending, they had a median of 437,571 views and 765 comments. This shows that trending videos get a lot of attention and interaction right away, even before they gather more views and comments over time.

**Categories and languages**. The most common categories were Entertainment (23M), People & Blogs (11.4M), and Music (11.2M), followed by Gaming (8.9M) and Sports (7.8M). Among videos with specified languages, the most frequent were English (7.6M), Arabic (2.9M), Spanish (1.9M), and Latin American Spanish (1.4M). These distributions reflect both platform-wide popularity patterns and regional linguistic structuring.

**Trend persistence**. Videos remained on the Trending list for a median of 217.14 hours (about nine days). Survival analysis shows that 90.83% of videos stayed for at least 48 hours, indicating that trending is not purely short-lived but often sustained.

**Cross-regional overlap**. A total of 187,742 videos appeared in two or more countries' Trending lists, with each video trending in an average of 2.26 countries. Only a small number of videos reached near-universal coverage across the 104 regions. Table 1 presents the top videos by global reach. Examples include Apple's September 7 Event in 2022 and MrBeast's *Every Country On Earth Fights For $250,000!*, both of which appeared in all covered regions, followed closely by BLACKPINK's *Pink Venom* music video. Such cases remain exceptional. As Figure 1 (log-scale) illustrates, the distribution of country coverage is highly skewed: most videos trended in only one or two countries, while only a few—typically major music releases, tech announcements, or large-scale stunts by global creators—achieve worldwide reach.

| # | Video Name | Channel (Video Genres) | Countries |
|---|---|---|---|
| 1 | Apple Event — September 7 | Apple (Science & Technology) | 104 |
| 2 | Every Country On Earth Fights For $250,000! | MrBeast (Entertainment) | 104 |
| 3 | BLACKPINK - 'Pink Venom' M/V | BLACKPINK (Music) | 104 |
| 4 | BLACKPINK - 'Shut Down' M/V | BLACKPINK (Music) | 103 |
| 5 | $1 Vs $100,000,000 Car! | MrBeast (Entertainment) | 103 |
| 6 | $1 vs $100,000,000 House! | MrBeast (Entertainment) | 103 |
| 7 | 정국 (Jung Kook)' Seven (feat. Latto)' Official MV | HYBE LABELS (Music) | 103 |
| 8 | Grand Theft Auto VI Trailer 1 | Rockstar Games (Gamig) | 103 |
| 9 | $1 vs $1,000,000,000 Yacht! | MrBeast (Entertainment) | 102 |
| 10 | 50 YouTubers Fight For $1,000,000 | MrBeast (Entertainment) | 102 |
| 11 | $1 vs $10,000,000 Job! | MrBeast (Entertainment) | 102 |

Table 1: Top videos by global reach

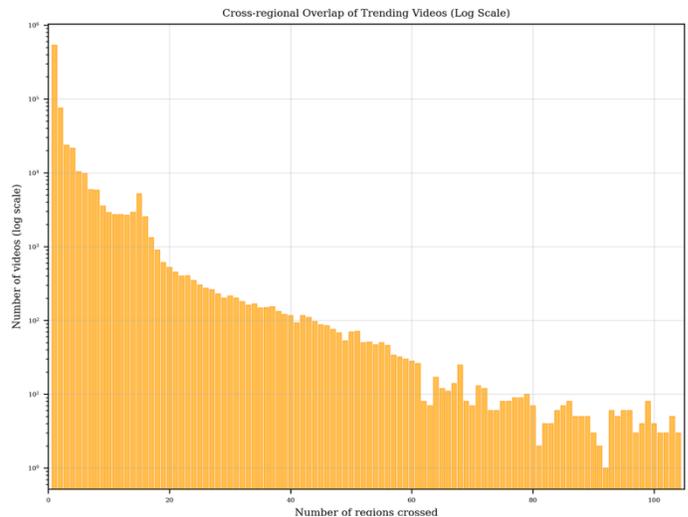

Figure 1: Log-scale distribution of videos by number of countries in which they trended

### 5.2 U.S.-Centric Trend Overview

**Sports dominate U.S. Trending.** Analysis of U.S. YouTube trending data showed that professional sports leagues and their media partners led the platform, headed by the NFL, NBA, and NBC Sports. Sports highlights and news drove most trending content. Although music labels such as HYBE (#8) and JYP (#14) and independent creators like Ryan Trahan (#19) appeared, the landscape was overwhelmingly shaped by institutional sports media rather than individual creators or traditional news outlets (see Table 2).

| # | Channel Name | Unique videos | # | Channel Name | Unique videos |
|---|---|---|---|---|---|
| 1 | NFL | 323 | 11 | WWE | 80 |
| 2 | NBA | 253 | 12 | ESPN FC | 79 |
| 3 | NBC Sports | 140 | 13 | Nintendo | 78 |
| 4 | CBS Sports Golazo | 124 | 14 | JYP Entertainment | 78 |
| 5 | FOX Soccer | 115 | 15 | Genshin Impact | 77 |
| 6 | FORMULA 1 | 104 | 16 | Fortnite | 77 |
| 7 | The United Stand | 98 | 17 | Netflix | 74 |
| 8 | HYBE LABELS | 96 | 18 | Clash of Clans | 73 |
| 9 | First We Feast | 83 | 19 | Ryan Trahan | 73 |
| 10 | ESPN | 83 | 20 | DAZN Boxing | 72 |

Table 2: Top channels in US Trending

**Gaming and entertainment dominate U.S. trending.** The genres that consistently dominated the U.S. trending videos list were Gaming (20.21%, 7,309 videos) and Entertainment (18.53%, 6,702 videos). Music (16.10%, 5,820 videos) and Sports (14.62%, 5,288 videos) also accounted for substantial shares of trending content. Conversely, many genres were nearly absent from trending status. Categories such as Movies, Anime/Animation, Horror, and Shorts each showed a 0.00% presence, indicating they rarely, if ever, achieved trending visibility. This pattern demonstrated YouTube's emphasis on broad, mass-appeal content, while niche genres remained largely excluded.

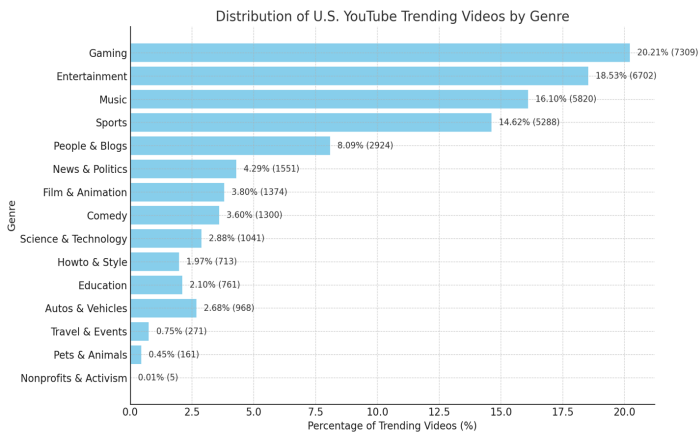

Figure 2: Percentage of US Trending videos by genre

*Note: Not a single video from cinematic genres like Movies, Horror, Documentary, or Anime, nor from niche categories like Shorts and Trailers, appeared on the U.S. Trending lists.*

**Longest U.S. Trending.** In the United States, MrBeast dominated the longest-trending list, with seven of the top ten videos each remaining on the charts for over 140 days. Sports content from DAZN Boxing and Red Bull also exhibited notable longevity, whereas music videos were absent from the top rankings. These patterns indicate that visibility beyond 100 days was concentrated among major independent creators and sports brands, while most videos trended for a relatively shorter time.

| # | Title | Channel (video genres) | Days | First Seen | Last Seen |
|---|---|---|---|---|---|
| 1 | Beat Ronaldo, Win $1,000,000 | MrBeast (Entertainment) | 147 | 2024-12-01 | 2025-01-06 |
| 2 | Every Country On Earth Fights For $250K | MrBeast (Entertainment) | 147 | 2023-08-20 | 2023-09-25 |
| 3 | Devin Haney vs. Ryan Garcia Highlights | DAZN Boxing (Sports) | 146 | 2024-04-21 | 2024-05-28 |
| 4 | $10,000 Every Day You Survive Grocery | MrBeast (Entertainment) | 146 | 2023-12-03 | 2024-01-08 |
| 5 | Escaping the LA Wildfires… | Royalty Family (People & Blogs) | 145 | 2025-01-20 | 2025-02-25 |
| 6 | I Spent 100 Hours Inside the Pyramids! | MrBeast (Entertainment) | 145 | 2025-02-09 | 2025-03-17 |
| 7 | Face Your Biggest Fear to Win $800,000 | MrBeast (Entertainment) | 144 | 2024-02-11 | 2024-03-18 |
| 8 | He Nearly Fell Off the Cliff 😬 | Red Bull (Sports) | 144 | 2024-12-02 | 2025-01-08 |
| 9 | I Survived 7 Days in an Abandoned City | MrBeast (Entertainment) | 144 | 2024-03-03 | 2024-04-08 |
| 10 | $1 vs. $100,000,000 House! | MrBeast (Entertainment) | 143 | 2023-10-15 | 2023-11-20 |

Table 3: Videos with the longest time on US Trending

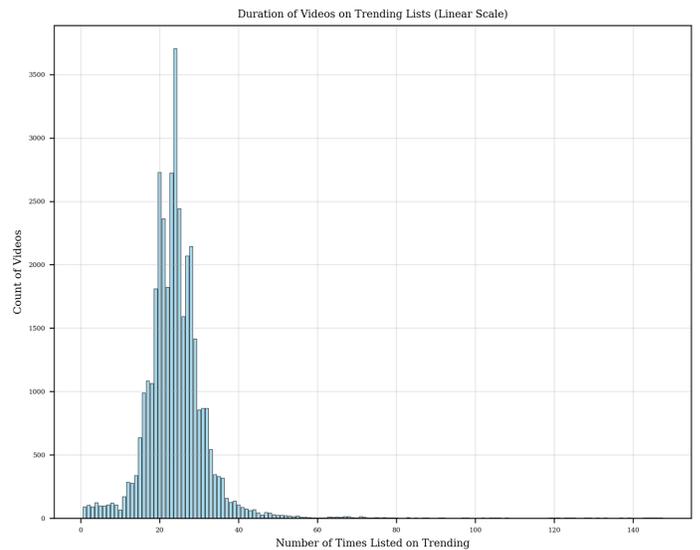

Figure 3: Distribution of duration of videos on the U.S. Trending list

**Event-driven content on U.S. Trending list**. To identify prominent social issues and events and examine their temporal patterns across multiple years, we conducted a detailed analysis of U.S. YouTube trending videos. We combined video titles and descriptions into a single text field and employed IBM Watson's Natural Language Understanding API to perform named-entity recognition, identifying people and institutions prominently featured in the trending content.

This analysis revealed four major event categories that warranted detailed tracking: U.S. politics, the Israel-Hamas conflict, the Russia-Ukraine war, and the British Royal Family. We then systematically identified trending videos on a weekly basis, flagging those containing at least one entity associated with these four categories, with each video counted only once per category. This method allowed us to compare attention patterns and competition among major social issues on YouTube's U.S. platform.

The distribution shows a consistently high baseline for U.S. politics, which intensifies during electoral cycles, as in the spikes of late 2024 and early 2025. This pattern reflects agenda-setting theory, where media emphasize certain issues and shape public priorities (McCombs and Shaw 1972). Domestic politics maintain salience because it connects to identity, civic stakes, and steady news output, whereas foreign or ceremonial topics appeared mainly through sharp, temporary surges triggered by external events.

The September 2022 rise in videos about the British Royals, linked to Queen Elizabeth II's death, illustrates how YouTube Trending curated ritualized, time-limited moments consistent with Dayan and Katz's (1992) "media events." Conflict coverage followed a different path: Israel-Hamas attention grew in October 2023 during escalation, while Russia-Ukraine peaked in early 2022 and remained elevated. These cases show how Trending captured global crises as cross-national cultural moments, with attention clustering at initiation or turning points and then leveling based on conflict intensity and content supply.

Together, these patterns highlight YouTube Trending as a finite and competitive attention space. U.S. political spikes often suppressed visibility for global issues, while quieter domestic periods opened room for international stories. The ranking system amplified these shifts: algorithms and creator responses generate feedback loops that expand content supply when demand rises. In this sense, YouTube Trending is not only a chart of popular videos but also a platform-curated lens on global affairs, shaping how digital publics engage with international news and producing shared narratives across national contexts.

## 6 Value of the Data

### 6.1 Published Work

Two published studies validate the dataset's value. First, in health communication, Ng (2023) conducted a cross-national analysis of fear appeals in YouTube Trending videos during the COVID-19 pandemic across the U.S., Brazil, Russia, Taiwan, Canada, and New Zealand. Using the Extended Parallel Process Model (EPPM) (Witte 1992; Witte and Allen 2000), the study examined how combinations of threat and efficacy messaging were framed across national contexts. Results showed that COVID-19 videos gained early attention in Taiwan but were delayed in the U.S. and Brazil, with Brazil featuring the fewest trending videos. Early exposure in Taiwan facilitated more effective public response, giving citizens time to adopt preventive measures and digest health information.

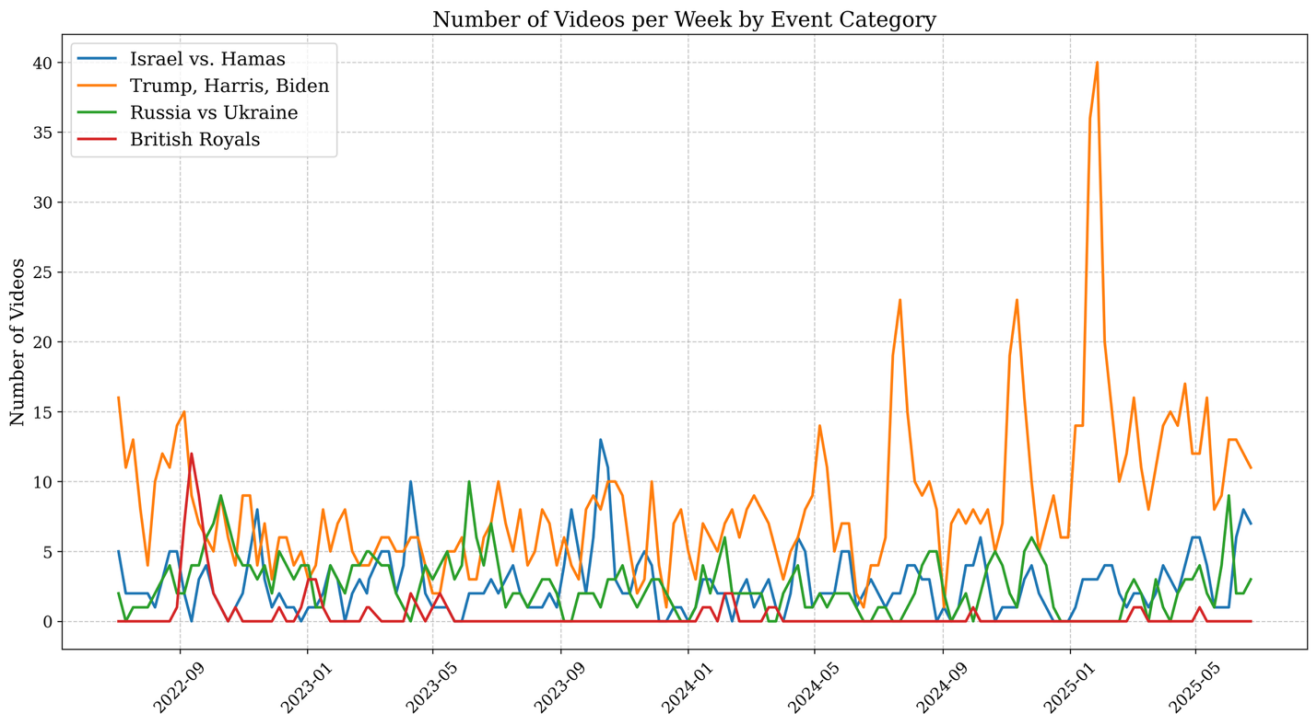

Figure 4: Event-Driven Content in the U.S. YouTube Trending, 2022-2025

Second, in the area of digital consumption, Ng and Taneja (2023) compared YouTube Trending data with Twitter trending topics and web traffic rankings. Their analysis showed that online media consumption remains strongly shaped by linguistic and geographic boundaries, challenging the notion of a borderless internet and revealing a "mosaic of regional cultures." Regional and national contexts continue to influence platform use and consumption patterns.

## 6.2 Other Possible Applications

The dataset opens up avenues for research that were previously unattainable. YouTube's rich, multimodal architecture—including video transcripts, metadata, and audience engagement metrics—enables large-scale content and sentiment analysis with an unparalleled level of granularity and scope. Its comprehensive coverage positions it as a vital resource for scholars investigating the dynamic interplay between platform design, user behavior, and global cultural flows.

**Cross-cultural media comparison**. How do digital trends traverse linguistic and national boundaries? With high-frequency captures, researchers can trace the trajectories of viral content across regions, enabling empirical tests of cultural proximity theory (La Pastina and Straubhaar 2005). This allows for a nuanced understanding of how cultural affinities and linguistic commonalities shape the spread of media content on global platforms.

**Crisis communication effectiveness**. The dataset offers actionable insights for NGOs, public health agencies, and policymakers. By analyzing framing patterns (e.g. humanitarian vs. blame-centered) across regions, stakeholders can assess which messages gain traction and how long they remain visible. For instance, during public health emergencies, the dataset can help identify which health messages rise to prominence, how quickly misinformation spreads compared to authoritative content, and which regions are most affected. A dashboard powered by daily Trending captures could serve as an early warning system, flagging spikes in content related to sensitive topics and enabling timely responses to disinformation campaigns.

**Platform governance and algorithmic accountability**. Trending data provides a rare, interface-level baseline for auditing platform behavior. Researchers can investigate whether YouTube's algorithm amplifies or suppresses content related to conflict, crisis, or dissent—and whether such patterns vary by region, language, or development level. What types of content are systematically favored or deprioritized? The dataset's scale and granularity support robust statistical analyses of algorithmic bias, offering empirical grounding for debates around platform governance, transparency, and accountability.

## 7 Ethical Considerations

### 7.1 Platform Compliance

This dataset comprises only publicly available metadata collected via the official YouTube Data API v3, in full compliance with the platform's Terms of Service (YouTube 2023). The data collection process was strictly limited to publicly accessible video and channel metadata; no personally identifiable information (PII) from individual user accounts—such as comments, user profiles, or watch histories—was gathered.

Researchers using this dataset are strongly encouraged to adhere to their institutional IRB (Institutional Review Board) protocols and respect platform-specific policies when conducting any enrichment tasks, such as retrieving transcripts or analyzing comment-level data. Ethical use of the dataset should prioritize user privacy, data minimization, and transparency in research practices.

### 7.2 FAIR Compliance

The dataset has been developed in accordance with the FAIR principles (Wilkinson et al., 2016), ensuring it is Findable, Accessible, Interoperable, and Reusable:

**Findable**. The dataset is assigned a persistent DOI and is accompanied by rich metadata to enhance discoverability through academic data repositories.

**Accessible**. The dataset is hosted on a trusted data-sharing platform supporting authorized access and standard retrieval protocols: https://databank.illinois.edu/policies.

**Interoperable**. Data are stored in standard CSV format, ensuring compatibility with widely used analysis tools and enabling integration with other datasets and workflows.

**Reusable**. Comprehensive documentation and metadata are provided to support interpretation and reuse across diverse research contexts. The dataset is distributed under terms permitting academic reuse, promoting transparency and reproducibility, with use restricted to research purposes only.

## 8 Limitations

### 8.1 YouTube Policy & API Changes

On December 13, 2021, YouTube modified its public API by removing the *dislikeCount* field from all video responses in line with updated platform policies (YouTube 2021). This change had unintended effects on our collection pipeline: because our retrieval script was structured around related engagement fields, the removal of *dislikeCount* also disrupted access to *likeCount*. As a result, our dataset includes *viewCount* and *commentCount* but not *likeCount* or *dislikeCount*.

## 8.2 API and Scheduling Constraints

Our three-year data collection achieved 98% coverage of targeted snapshots (446,971 of 455,936), with brief outages accounting for the small gaps. The high-frequency sampling supports analysis of short-term attention spikes and temporal patterns, though the missing snapshots may limit studies requiring continuous temporal coverage. While relatively minor in proportion, these gaps may affect the completeness of time-sensitive or event-driven analyses and should be acknowledged in studies requiring continuous temporal coverage.

# 9 Conclusion

The retirement of YouTube's Trending page on July 1, 2025, marks a significant turning point in in documenting global video culture. The transition from a platform-curated, non-personalized discovery interface to a fully algorithmic, hyper-personalized model fundamentally changed how content is surfaced, circulated, and consumed worldwide.

This three-year archival dataset preserves a now-defunct layer of platform infrastructure and serves as a rare empirical record of global popular culture in the algorithmic age. Its breadth and granularity offer valuable opportunity for researchers to detect subtle patterns, rare events, and long-term shifts in content popularity. As such, it stands as a unique and irreplaceable historical artifact, and a foundational resource for research in platform studies, computational social science, digital communication, and media policy.

# Acknowledgements

This work was supported by the Campus Research Board and the Institute of Communications Research at the University of Illinois. The authors utilized GPT-5 and Claude-Opus-4 to assist with research, table and citation formatting, and manuscript readability. The authors reviewed and approved all final content.

# Ethics Checklist

1. For most authors…

(a) Would answering this research question advance science without violating social contracts, such as violating privacy norms, perpetuating unfair profiling, exacerbating the socio-economic divide, or implying disrespect to societies or cultures? Yes. The research uses only publicly available metadata via the YouTube Data API v3. No personally identifiable information is collected.

(b) Do your main claims in the abstract and introduction accurately reflect the paper's contributions and scope? Yes. The abstract and introduction accurately describe the dataset

(three years, 104 countries, 446,971 snapshots) and its contributions (longitudinal, cross-national, non-personalized visibility data).

- (c) Do you clarify how the proposed methodological approach is appropriate for the claims made? Yes. The methodological approach (API collection, four snapshots/day, cross-country coverage) is clearly justified for claims about temporal patterns, attention spikes, and cross-national comparisons.
- (d) Do you clarify what are possible artifacts in the data used, given population-specific distributions? Yes. We noted potential artifacts: skewed distributions (e.g., most videos trend in only 1–2 countries) and concentrated visibility among global stars.
- (e) Did you describe the limitations of your work? Yes. We explicitly described limitations: API changes (like removal of dislikeCount/likeCount) and minor gaps in snapshot coverage.
- (f) Did you discuss any potential negative societal impacts of your work? Yes. We mentioned ethical considerations and encouraged responsible use, including adherence to IRB protocols. While societal harm is minimal, they acknowledge potential indirect effects (e.g., analyzing controversial content).
- (g) Did you discuss any potential misuse of your work? Yes. We specified that researchers have access to additional data such as complete videos (both visual and audio), transcripts, and viewer comments. We also address data use in the Ethical and FAIR Considerations section, noting that the dataset is restricted to research purposes only.
- (h) Did you describe steps taken to prevent or mitigate potential negative outcomes of the research, such as data and model documentation, data anonymization, responsible release, access control, and the reproducibility of findings? Yes. Steps taken include only collecting public metadata, anonymization by design (no user accounts), responsible release, and documentation to support reproducibility.
- (i) Have you read the ethics review guidelines and ensured that your paper conforms to them? Yes. The paper aligns with standard ethics guidelines, including IRB considerations and YouTube TOS compliance.

2. Additionally, if your study involves hypotheses testing…
   (a) Did you clearly state the assumptions underlying all theoretical results? Not applicable
   (b) Have you provided justifications for all theoretical results? Not applicable
   (c) Did you discuss competing hypotheses or theories that might challenge or complement your theoretical results? Not applicable
   (d) Have you considered alternative mechanisms or explanations that might account for the same outcomes observed in your study? Not applicable
   (e) Did you address potential biases or limitations in your theoretical framework? Not applicable
   (f) Have you related your theoretical results to the existing literature in social science? Not applicable
   (g) Did you discuss the implications of your theoretical results for policy, practice, or further research in the social science domain? Not applicable

3. Additionally, if you are including theoretical proofs…
   (a) Did you state the full set of assumptions of all theoretical results? Not applicable
   (b) Did you include complete proofs of all theoretical results? Not applicable

4. Additionally, if you ran machine learning experiments…
   (a) Did you include the code, data, and instructions needed to reproduce the main experimental results (either in the supplemental material or as a URL)? Not applicable
   (b) Did you specify all the training details (e.g., data splits, hyperparameters, how they were chosen)? Not applicable
   (c) Did you report error bars (e.g., with respect to the random seed after running experiments multiple times)? Not applicable
   (d) Did you include the total amount of compute and the type of resources used (e.g., type of GPUs, internal cluster, or cloud provider)? Not applicable
   (e) Do you justify how the proposed evaluation is sufficient and appropriate to the claims made? Not applicable
   (f) Do you discuss what is "the cost" of misclassification and fault (in)tolerance? Not applicable

5. Additionally, if you are using existing assets (e.g., code, data, models) or curating/releasing new assets, without compromising anonymity…
   (a) If your work uses existing assets, did you cite the creators? Yes. The dataset uses the official YouTube API, and prior work is cited throughout.
   (b) Did you mention the license of the assets? Yes. YouTube Data API usage is under standard license; this is noted in Ethical Considerations.
   (c) Did you include any new assets in the supplemental material or as a URL? The dataset is released via DOI: https://doi.org/10.13012/B2IDB-9307654_V1
   (d) Did you discuss whether and how consent was obtained from people whose data you're using/curating? Yes. Consent is implicitly addressed: only public metadata is collected, consistent with TOS.
   (e) Did you discuss whether the data you are using/curating contains personally identifiable information or offensive content? Yes. No PII is included. Offensive content is not collected systematically, though videos may contain it; the paper notes this risk indirectly.
   (f) If you are curating or releasing new datasets, did you discuss how you intend to make your datasets FAIR (see FORCE11 (2020))? Yes. FAIR considerations discussion is provided.
   (g) If you are curating or releasing new datasets, did you create a Datasheet for the Dataset (see Gebru et al. (2021))? Yes. We provided comprehensive metadata and documentation.

6. Additionally, if you used crowdsourcing or conducted research with human subjects, without compromising anonymity…
   (a) Did you include the full text of instructions given to participants and screenshots? Not applicable
   (b) Did you describe any potential participant risks, with mentions of Institutional Review Board (IRB) approvals? Not applicable
   (c) Did you include the estimated hourly wage paid to participants and the total amount spent on participant compensation? Not applicable
   (d) Did you discuss how data is stored, shared, and de-identified? Not applicable

# Appendix

The dataset contains 104 countries.

**Africa** (13 countries)
Algeria (DZ), Egypt (EG), Ghana (GH), Kenya (KE), Libya (LY), Morocco (MA), Nigeria (NG), Senegal (SN), Tunisia (TN), Tanzania (TZ), Uganda (UG), South Africa (ZA), and Zimbabwe (ZW).

**Asia** (30 countries)
United Arab Emirates (AE), Azerbaijan (AZ), Bahrain (BH), Cyprus (CY), Georgia (GE), Hong Kong (HK), Indonesia (ID), Israel (IL), India (IN), Iraq (IQ), Jordan (JO), Japan (JP), South Korea (KR), Kuwait (KW), Kazakhstan (KZ), Lebanon (LB), Sri Lanka (LK), Malaysia (MY), Nepal (NP), Oman (OM), Philippines (PH), Pakistan (PK), Qatar (QA), Saudi Arabia (SA), Singapore (SG), Thailand (TH), Turkey (TR), Taiwan (TW), Vietnam (VN), and Yemen (YE).

**Europe** (38 countries)
Austria (AT), Bosnia and Herzegovina (BA), Belgium (BE), Bulgaria (BG), Belarus (BY), Switzerland (CH), Czech Republic (CZ), Germany (DE), Denmark (DK), Estonia (EE), Spain (ES), Finland (FI), France (FR), United Kingdom (GB), Greece (GR), Croatia (HR), Hungary (HU), Ireland (IE), Iceland (IS), Italy (IT), Liechtenstein (LI), Lithuania (LT), Luxembourg (LU), Latvia (LV), Montenegro (ME), North Macedonia (MK), Malta (MT), Netherlands (NL), Norway (NO), Poland (PL), Portugal (PT), Romania (RO), Serbia (RS), Russia (RU), Sweden (SE), Slovenia (SI), Slovakia (SK), and Ukraine (UA).

**Latin America and the Caribbean** (19 countries)
Argentina (AR), Bolivia (BO), Brazil (BR), Chile (CL), Colombia (CO), Costa Rica (CR), Dominican Republic (DO), Ecuador (EC), Guatemala (GT), Honduras (HN), Jamaica (JM), Mexico (MX), Nicaragua (NI), Panama (PA), Peru (PE), Puerto Rico (PR), Paraguay (PY), El Salvador (SV), and Uruguay (UY).

**Northern America (2** countries**)**
Canada (CA) and United States (US).

**Oceania (2 countries)**
Australia (AU) and New Zealand (NZ).